\documentclass[aps,pre,twocolumn]{revtex4-1}
\usepackage{amssymb}
\usepackage{bm}
\usepackage{url}
\usepackage{graphicx}

\usepackage[dvipsnames,usenames]{color}
\begin{document}

\title{Split energy cascade in turbulent thin fluid layers}

\author{Stefano Musacchio}
\affiliation{Universit\'e C\^ote d'Azur, CNRS, LJAD, Nice, France}

\author{Guido Boffetta}
\affiliation{Dipartimento di Fisica and INFN, Universit\`a di Torino, 
via P. Giuria 1, 10125 Torino, Italy}

\begin{abstract}
We discuss the phenomenology of the split energy cascade in a 
three-dimensional thin fluid layer by mean of high resolution
numerical simulations of the Navier-Stokes equations. 
We observe the presence of both an inverse energy cascade at large
scales, as predicted for two-dimensional turbulence, and of a direct energy
cascade at small scales, as in three-dimensional turbulence.
The inverse energy cascade is associated with a direct cascade of
enstrophy in the intermediate range of scales. 
Notably, we find that the inverse cascade of energy in this system
is not a pure 2D phenomenon, as the coupling with the
3D velocity field is necessary to guarantee the constancy of fluxes.
\end{abstract}


\maketitle 

\section{Introduction}
\label{sec:1}

Fifty years ago, Kraichnan showed in the seminal paper
\cite{kraichnan1967inertial}
that the dynamics of an incompressible flow in
two-dimensions (2D) is dramatically different from the classical
phenomenology of three-dimensional (3D) turbulence. 
The presence of two inviscid quadratic invariants, energy and enstrophy, 
gives rise to a double-cascade scenario \cite{eyink1996exact}.
At variance with the 3D case, in which the kinetic energy 
cascades toward small viscous scales,  
in 2D it is transferred toward large scales. Such ``inverse energy
cascade'' is accompanied by a ``direct enstrophy cascade'', which
proceeds towards small scales \cite{boffetta2012two}.
In the inverse and direct ranges of scales the theory predicts 
a kinetic energy spectrum $E(k) \simeq \varepsilon_{f}^{2/3} k^{-5/3}$
and $E(k) \simeq \eta_{f}^{2/3} k^{-3}$ with possible logarithmic
corrections \cite{kraichnan1971inertial}. Here and in the following
$\varepsilon_{f}$ and $\eta_{f}$ denote the energy and the enstrophy
injection rates respectively.
The Kraichnan seminal concept of inverse cascade has become
since then a prototypical model for several turbulent systems,
from the inverse cascade in strongly rotating 3D flows 
\cite{pouquet2013inverse}, to the inverse cascade of magnetic helicity 
in three-dimensional
magneto-hydrodynamic turbulence \cite{frisch1975possibility}, of
passive scalar in compressible turbulence \cite{chertkov1998inverse},
of wave action in weak turbulence \cite{zakharov1982kinetic}.

The presence of the two cascades in two-dimensional turbulence 
has been observed in a number of numerical simulations 
\cite{frisch1984numerical,herring1985comparison,maltrud1991energy,smith1993bose,boffetta2000inverse,boffetta2002intermittency,chen2003physical,xiao2009physical,boffetta2010evidence,cencini2011nonlinear}
and in experiments in soap films
\cite{rutgers1998forced,belmonte1999velocity,rivera1998turbulence,rivera2000external,rivera2002homogeneity,bruneau2005experiments,rivera2014direct}
and in thin fluid layers
\cite{paret1997experimental,paret1999vorticity,kellay2002two}. 

At variance with the numerical investigations, which allows to 
to study the ideal 2D Navier-Stokes equations, 
the experiments have to deal with the effects of the finite thickness
of the fluid layer. 
This issue, which has been often considered a limitation for two-dimensional
experiments,  opens a series of interesting questions.  
How is the ideal 2D phenomenology modified in the case of a thin (3D) fluid
layer?  
How thin should the layer be, in order to display the 2D-like double cascade? 
In which way does the transition from the 2D to the 3D regime occur at
increasing the thickness of the layer? 

These questions have been addressed both in numerical simulations
\cite{smith1996crossover,celani2010turbulence}
and experiments \cite{xia2011upscale,byrne2011robust}
which have shown that the critical parameter which controls the cascade
direction is the ratio $S=L_z/L_f$ of the thickness to the forcing
scale. When $S \ge 1$ the flow is 3D at the forcing scale and the 
injected energy produces a direct cascade as in usual 3D turbulence.
By decreasing $S$ one observes the phenomenon of {\it cascade splitting}
with a fraction of the injected energy which goes to large scales
and the remaining energy which flows to small scales 
\cite{smith1996crossover,celani2010turbulence}.
For $S \ll 1$ the flux of the direct energy cascade is expected to scale as $S^2$. 
This prediction has been verified in shell models for quasi-two-dimensional turbulence
\cite{boffetta2011shell}. 
Further reducing $S$, the cascade of energy towards small scale vanishes 
when the thickness $L_z$ reaches the Kolmogorov scale $L_\nu$ 
and the flow recovers the standard 2D phenomenology.

In the present paper we investigate the entanglement 
of 2D and 3D dynamics which occurs in a turbulent fluid layer
by means of numerical simulations of the 3D Navier-Stokes equations
in a confined domain with $L_z < L_f$ (and $L_z > L_\nu$).
In agreement with previous findings, we observe the phenomenon 
of splitting of the energy cascade. 
By introducing a suitable decomposition of the velocity field,
we show that the inverse cascade 
involves mainly the kinetic energy of the 2D modes, 
while the energy of the remnant 3D velocity is transferred 
toward the viscous scales. 
We also show that the development of the inverse energy cascade is
associated with a partial conservation of the enstrophy 
in the intermediate range of scales $L_z < \ell < L_f$. 
Interestingly, we find that 3D modes 
play a relevant role in the the 2D phenomenology 
which is observed at large scales. 
In particular, the transport of the 2D modes by the 3D velocity 
is necessary to ensure a constant flux of energy in the inverse cascade
as well as a constant flux of enstrophy in the range $L_z < \ell < L_f$.

The remaining of the paper is organized as follows. 
Section~\ref{sec:2} introduces the Navier-Stokes equations 
and the decomposition of the velocity field in the 2D and 3D modes. 
In Section~\ref{sec:3} we report the results of the numerical simulations. 
Section~\ref{sec:4} is devoted to the conclusions. 
In the Appendix~\ref{appA} we derive a 2D model 
for the dynamics of a thin layer in the limit $L_z \to 0$. 

\section{Navier-Stokes equations for a thin layer}
\label{sec:2}

We consider the dynamics of a three-dimensional thin layer of fluid, 
ruled by the
Navier-Stokes equations for the velocity field ${\bm u}({\bm x},t)$: 
\begin{equation} 
\label{eq:1}
\partial_t {\bm u} + {\bm u} \cdot {\bm \nabla} {\bm u} = 
-{\bm \nabla} p + \nu \Delta {\bm u} + {\bm f} 
\end{equation} 
where $\nu$ is the kinematic viscosity,  
${\bm f}$ is the external forcing, and the pressure $p$ is
determined by the incompressibility constraint
${\bm \nabla} \cdot {\bm  u} = 0$. 
The flow is confined in a thin domain of size $L_x=L_y= r L_z$ 
with $r \gg 1$ and periodic boundary conditions in all the 
directions.

In absence of forcing and dissipation, 
(\ref{eq:1}) preserves the kinetic energy 
$E = (1/2) \langle |{\bm  u}|^2 \rangle$ (where 
$\langle ... \rangle$ denotes average over the space).
The energy balance in the forced-dissipated case reads: 
\begin{equation} 
\label{eq:2} 
\frac{dE}{dt} = \varepsilon_f - \varepsilon_\nu
\end{equation} 
where 
$\varepsilon_\nu = \nu \langle ({\bm \nabla} {\bm u})^2 \rangle$ 
is the energy dissipation rate due to the viscosity and 
$\varepsilon_f = \langle {\bm f} \cdot {\bm u} \rangle $ 
is the energy input provided by the external forcing. 

The forcing provides also an input of enstrophy 
$Z = (1/2) \langle |{\bm  \omega}|^2 \rangle$, 
(${\bm \omega} = {\bm \nabla} \times {\bm u}$ denotes
the vorticity field)
at the rate 
$\eta_f = \langle ({\bm \nabla} \times{\bm f}) \cdot {\bm \omega}
\rangle$. 
At variance with the ideal 2D case, the enstrophy is not preserved by
the full 3D inviscid dynamics, but it is produced by the vortex stretching 
mechanism \cite{tennekes1972first}. The equation for the vorticity
field is obtained by taking the curl of (\ref{eq:1})
\begin{equation} 
\label{eq:3}
\partial_t {\bm \omega} + {\bm u} \cdot {\bm \nabla} {\bm \omega} = 
 {\bm \omega} \cdot {\bm \nabla} {\bm u} + \nu \Delta {\bm \omega} + {\bm f_\omega} 
\end{equation} 
where ${\bm f_\omega} = {\bm \nabla} \times{\bm f}$ and
${\bm \omega} \cdot {\bm \nabla} {\bm u}$ represents the 
vortex stretching term.

In order to highlight the presence of a 2D phenomenology
in the 3D flow, it is useful to decompose the 
velocity field as ${\bm u} = {\bm  u}^{2D} + {\bm u}^{3D}$. 
The 2D mode ${\bm  u}^{2D}=(u_x^{2D}(x,y),u_y^{2D}(x,y),0)$ 
is defined as the average along the $z$ direction of the
$x$ and $y$ components of the velocity field ${\bm u}$, 
and it satisfies the 2D incompressibility condition
$\partial_x u_x^{2D} + \partial_y u_y^{2D} = 0$. 
In the Fourier space it corresponds to the mode $k_3=0$ of the
horizontal velocity 
${\bm  u}^{2D}(k_1,k_2) = ( u_x(k_1,k_2),
u_y(k_1,k_2),0)$. 
The field ${\bm u}^{3D}$ is defined as the difference
${\bm u}^{3D}={\bm u}-{\bm  u}^{2D}$.
By the above definitions it is easy to show that the 
total energy decomposes into a 2D contribution 
$E^{2D}=(1/2) \langle |{\bm u}^{2D}|^2 \rangle$ and 
a 3D contribution
$E^{3D}=(1/2) \langle |{\bm u}^{3D}|^2 \rangle$ 
as $E=E^{2D}+E^{3D}$.
We notice that, beside the kinetic energy of the vertical velocity, 
$E^{3D}$ contains also the contributions of the modes $k_3 \neq 0$ 
of the horizontal components of the velocity.

Similarly, the vorticity field can be decomposed as 
${\bm \omega} = \omega^{2D} \hat{\bm z} + {\bm \omega}^{3D}$, 
where $\omega^{2D} = \partial_x u_y^{2D} - \partial_y u_x^{2D}$
is the scalar vorticity of the two-dimensional flow.
In the limit of vanishing thickness $L_z \to 0$ 
(at finite viscosity $\nu$) the vertical dependence disappears
and ${\bm  u}^{2D}$ becomes solution of the 2D Navier-Stokes equation
(see Appendix~\ref{appA}). 
Therefore, it is reasonable to assume that the occurrence of an
inverse energy cascade at finite thickness $L_z$ should be
associated to the dynamics of the 2D mode ${\bm  u}^{2D}$. 

\section{Direct numerical simulations}
\label{sec:3}

We performed a direct numerical simulation
of the Navier-Stokes equations (\ref{eq:1}) in a confined geometry
with periodic boundary conditions.
The computational domain has dimensions 
$L_x=L_y=2\pi$, $L_z = L_x/64$ ($r=64$) and it is 
discretized on a uniform grid at resolution 
$N_x \times N_y\times N_z = 4096 \times 4096 \times 64$. 
The numerical simulations are performed by means of a 
fully-parallel, pseudospectral code, with $2/3$ dealiasing
scheme.   
We adopt an hyperviscous 
damping scheme $(-1)^{p-1} \nu_p \Delta^p$
with $p=4$ and  $\nu_p = 10^{-21}$. 
We do not use any large-scale dissipation (such as linear friction).

\begin{figure}[h!]
\centering
\includegraphics[width=0.8\columnwidth]{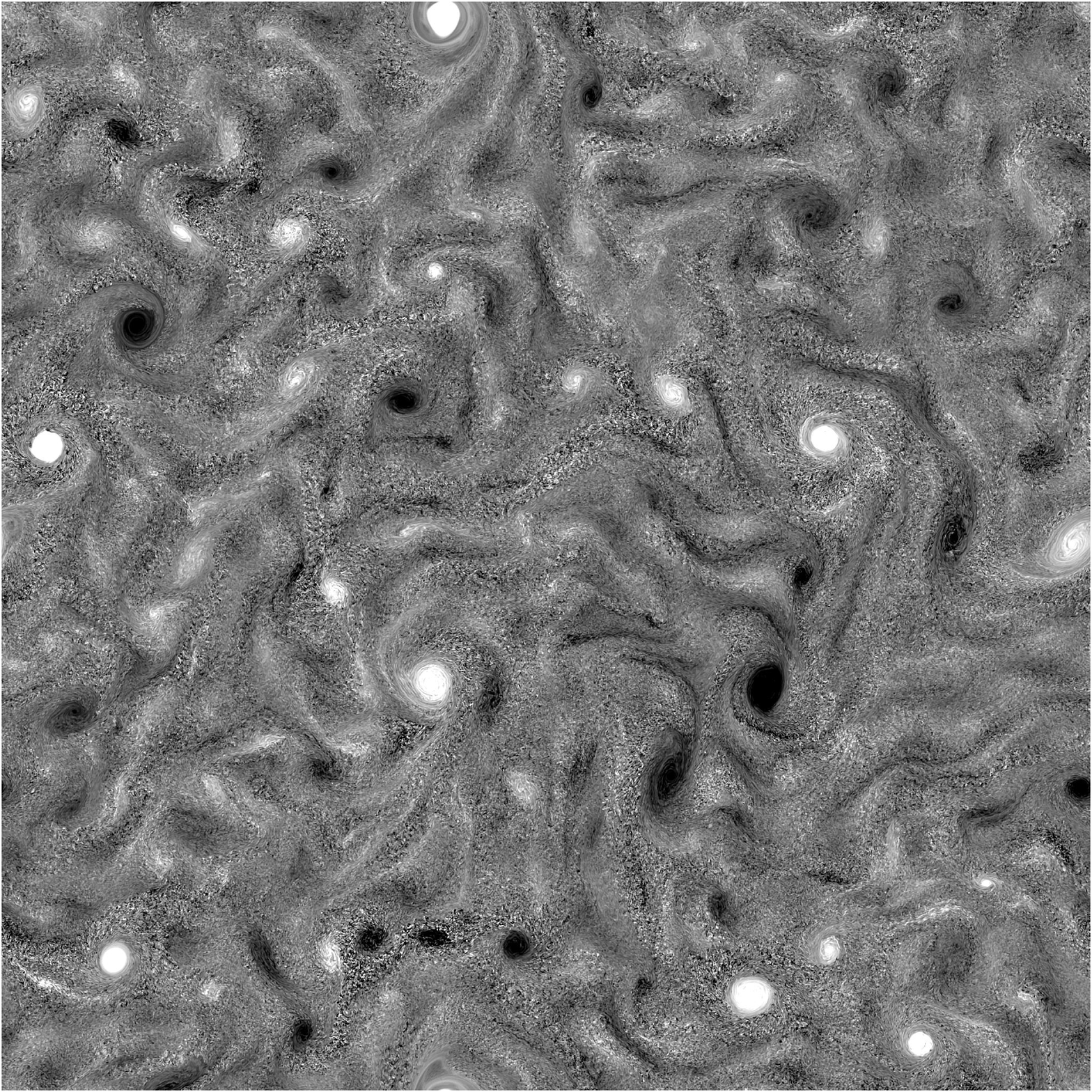}
\caption{Snapshot of the scalar vorticity field $\omega^{2D}$
in the late stage of the simulation. Typical two-dimensional objects,
such as strong vortices at the forcing scale $L_f$, coexist 
with small scale three-dimensional features.}
\label{fig:1}
\end{figure}

The flow is sustained by a 
``two-components, two-dimensional'' forcing, that is,  
the forcing is active on the horizontal components of the velocity 
and it is dependent on the horizontal coordinates only, 
${\bm f} = (f_x(x,y),f_y(x,y),0)$. Therefore in the vorticity
equation (\ref{eq:3}) forcing ${\bm f}_{\omega}$ is active on 
the 2D field $\omega^{2D}$ only.
Forcing is restricted to a narrow wavenumber shell in Fourier
space with 
$k_h = (k_1^2+k_2^2)^{1/2} \simeq k_f , k_3=0$. Here $k_f = 16$. 
The forcing is Gaussian and $\delta$-correlated in time to control 
the injection rates $\varepsilon_f$ and $\eta_f$.
The characteristic time at the forcing scale is defined as
$\tau_f = \eta_f^{-1/3}$.  
The ratio between the thickness $L_z= 2 \pi /k_z$ and the forcing scale
$L_f = 2 \pi/k_f$ is $S=L_z/L_f = 4$ 
This ensures the regime of split cascade with the 
coexistence of the 3D and the 2D
phenomenology (see Fig.~\ref{fig:1}) \cite{celani2010turbulence}.

\begin{figure}[h!]
\centering
\includegraphics[width=1.0\columnwidth]{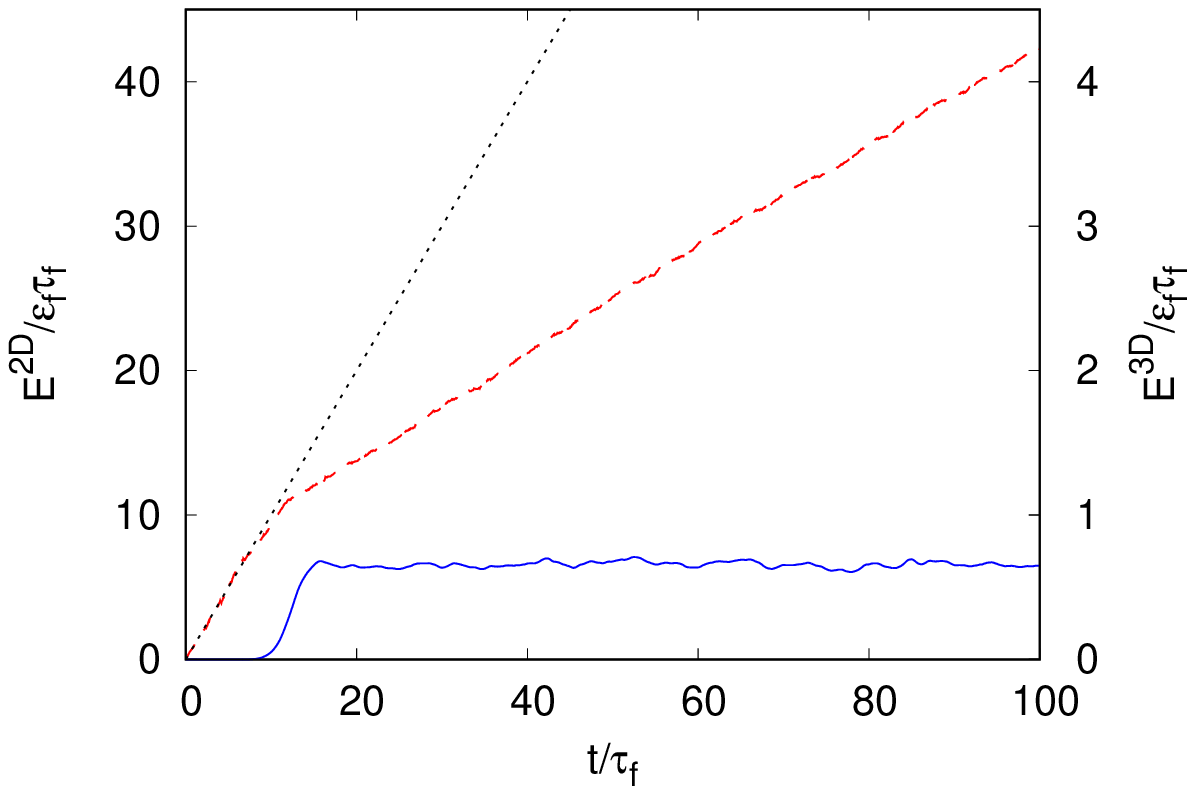}
\caption{Temporal evolution of the kinetic energy 
$E^{2D}$ (red, dashed line, left $y$-axis) and  
$E^{3D}$ (blue, solid line, right $y$-axis). 
The dotted line represents the linear growth with 
the input rate $\varepsilon_f$ (left $y$-axis).}
\label{fig:2}
\end{figure}

The velocity field at time $t=0$ is initialized to zero, plus 
a small random perturbation, which is required to trigger the 3D
instability. The results shown in this Section has been obtained
with an energy of the initial perturbation 
$E_{pert} \simeq 1.6 \cdot 10^{-7} \varepsilon_f \tau_f$.  

Because of the purely 2D forcing,  
in the early stage of the simulation ($t < 10 \tau_f$), 
the energy accumulates in the 2D mode ${\bm u}^{2D}$ at a rate equal
to the forcing input, while the energy $E^{3D}$ of the 3D component is
negligible (see Figure~\ref{fig:2}). 
The activation of the 3D modes ${\bm u}^{3D}$ (at $t \simeq 10 \tau_f$)
is accompanied by a significant reduction of the growth rate of the 2D
energy as some of the injected energy is now transferred to small scales.
At later times, $E^{3D}$ saturates to a statistically
steady value (as in standard 3D turbulence), while the two-dimensional
component $E^{2D}$ energy keeps increasing with a constant reduced rate
$\varepsilon_\alpha < \varepsilon_f$. 
We stop the simulation at time $t= 100 \tau_f$, when the inverse energy cascade 
has reached the lowest wavenumber (see Figure~\ref{fig:4}). 
Continuing the simulation for further times, 
in absence of large-scale dissipation, we expect that 
the kinetic energy will accumulate in the lowest mode,
giving rise to the so called ``condensate''
~\cite{chertkov2007dynamics,xia2009spectrally,laurie2014universal}. 
The 2D mode contains almost all the kinetic energy of the horizontal
velocities, and the kinetic energy of the vertical component $u_z$
contained in the mode $k_3=0$ represents only $24\%$ of the total. 

\begin{figure}[h!]
\centering
\includegraphics[width=1.0\columnwidth]{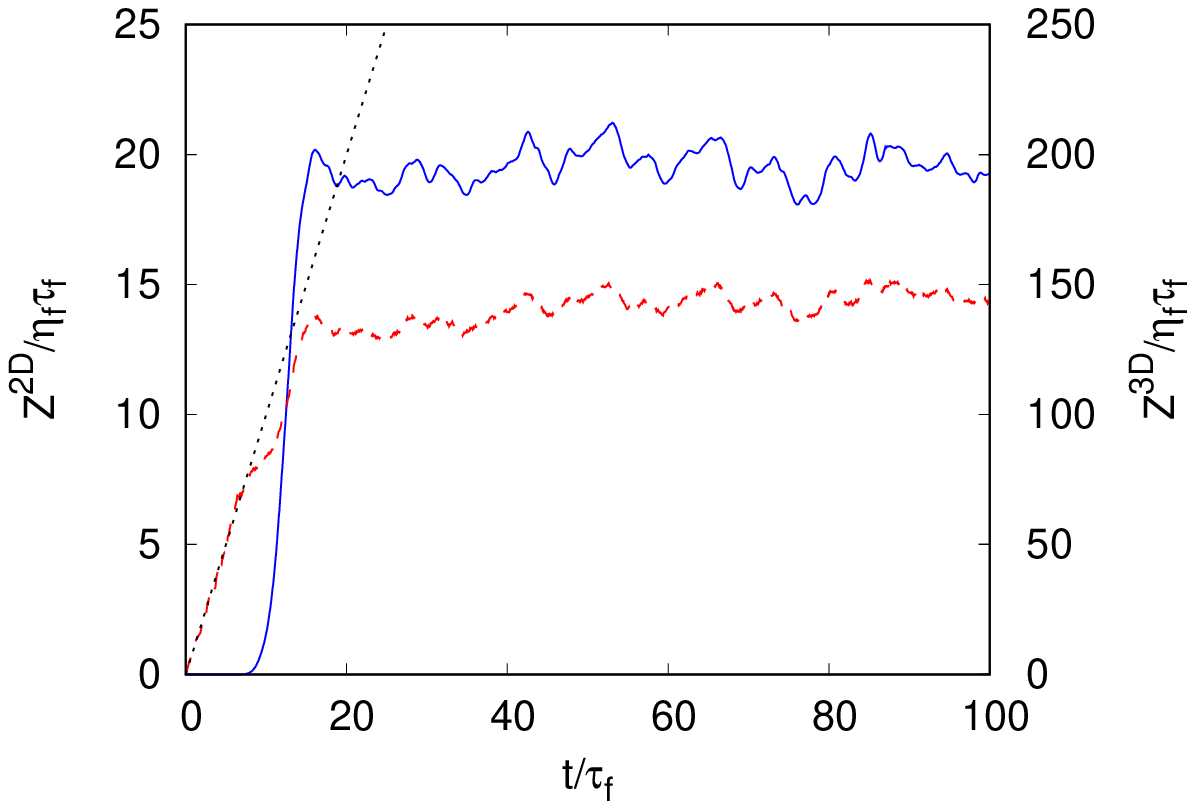}
\caption{Temporal evolution of the enstrophy of 
$\omega^{2D}$ (red, dashed line, left $y$-axis) and  
${\bm \omega}^{3D}$ (blue, solid line, right $y$-axis). 
The dotted line represents the linear growth with 
the input rate $\eta_f$ (left $y$-axis).}
\label{fig:3}
\end{figure}
The enstrophy of the 2D mode 
$Z^{2D}= 1/2 \langle  (\omega^{2D})^2 \rangle$ 
grows initially with the input rate $\eta_f$, reflecting the pure 
2D nature of the initial flow (see Figure~\ref{fig:3}). 
As 3D motions develop (at 
$t \simeq 10 \tau_{f}$), the enstrophy
associated to the 3D modes 
$Z^{3D}= 1/2 \langle  |{\bm \omega}^{3D}|^2 \rangle$
increases very rapidly and reaches a stationary value which is 
much larger than the saturation level of $Z^{2D}$.

\subsection{Energy spectra}

\begin{figure}[h!]
\centering
\includegraphics[width=1.0\columnwidth]{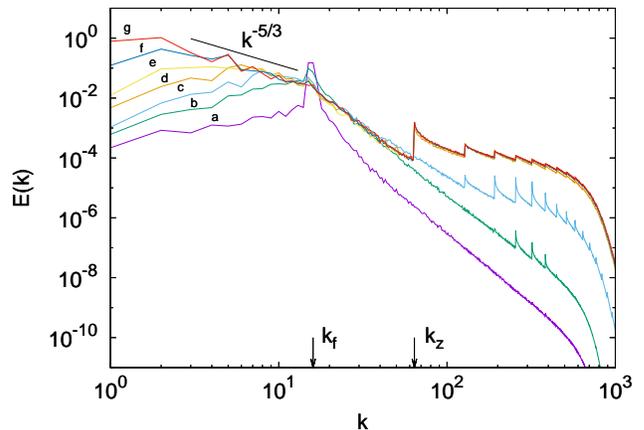}
\caption{
Energy spectra $E(k,t)$ at times 
$t/\tau_f =$ 
$4.6$ (a, violet), 
$6.9$ (b, green), 
$9.2$ (c, cyan), 
$13.7$ (d, orange), 
$18.3$ (e, yellow), 
$45.8$ (f, blue), 
$91.6$ (g, red).
The last three spectra are almost superposed for $k > k_z$. 
}
\label{fig:4}
\end{figure}
In Figure~\ref{fig:4} we show the instantaneous spectra $E(k,t)$ 
of the total energy at different times $t$. 
The initial spectra are almost completely 2D, because the forcing is
active only on 2D modes. 
We observe that the 3D instability begins 
at high harmonics of the thickness wavenumber $k_z$, 
then it propagates to all the modes $k > k_z$.  
The energy spectrum at $k > k_f$ 
saturates at time $t \simeq 16
\tau_f$. 
At scales larger than the forcing scale, $k < k_f$,
 we observe the development of an
inverse energy cascade with a power-law spectrum $E(k) \simeq
\varepsilon_{\alpha}^{2/3} k^{-5/3}$. 

The ``spiky'' aspect of the energy spectrum at high wavenumbers $k >
k_z$ is due to the anisotropic spacing of the wavenumbers in
the Fourier space. The separation between the discrete wavenumbers in the
horizontal direction is $\Delta k_h = 2 \pi /L_x$ while in the vertical
direction it is $\Delta k_3 = 2 \pi  / L_z = k_z $. 
Given that $L_z \ll L_x$,  the wavenumber space is structured as
horizontal dense layers, separated by large gaps in the vertical
direction. 
Because the complete energy spectrum is defined as the integral of 
the square amplitude
of the modes over a spherical wavenumber shell of radius $k$, one gets
a sudden increase of the spectrum each time the spherical shell is a
multiple of $k_z$. 

In order to analyze the contribution of the 2D mode to the total
energy spectrum, it is interesting also to consider the 2D spectra 
$E^{2D}(k)$, in which the integral is restricted to the horizontal
wavenumbers $k_h = \sqrt{k_1^2 +k_2^2}$ on the plane $k_3=0$. 
In physical space, this is equivalent of averaging first the velocity
fields in the vertical direction $z$ and then computing the spectrum
of the averaged 2D fields. 
It is worth to notice that, for $k< k_z$, the 2D spectra and 3D
spectra coincide, because the spherical shell of radius $k < k_z$ 
intersects the planes $k_3= \pm m k_z$ only for $m=0$. 

\begin{figure}[h!]
\centering
\includegraphics[width=1.0\columnwidth]{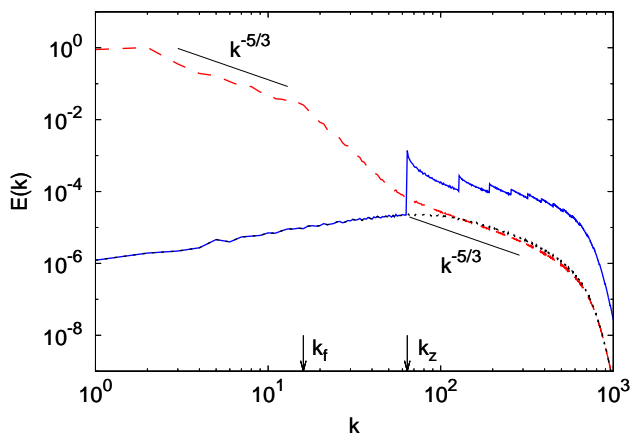}
\caption{Energy spectrum $E^{2D}(k)$ of the 
2D mode ${\bm u}^{2D}$ (red, dashed line) and
3D mode ${\bm u}^{3D}$ (blue, solid line). 
We also show the 2D spectrum of the 
vertically-averaged vertical velocity $u_z$ (black, dotted line).  
The spectra are computed at $t = 100 \tau_f$}
\label{fig:5}
\end{figure}
In Figure~\ref{fig:5} we compare the 2D energy spectrum $E^{2D}(k)$
of the 2D mode  ${\bm u}^{2D}$ with the 3D energy spectrum of the 
3D mode ${\bm u}^{3D}$. 
At low wavenumbers $k < k_z$ almost all the kinetic energy is
contained in the 2D mode. 
This confirms that the inverse cascade 
which develops in the range $k < k_f$ concerns only the 2D energy. 
Conversely, the 3D mode contains the largest fraction 
of the kinetic energy at high wavenumbers $k> k_z$. 
Interestingly, in the same range of wavenumber, 
the 2D spectrum of the 2D mode displays a $-5/3$ slope, 
and it is very close to the 2D spectrum 
of the vertical component $u_z$. 

The spectrum $E^{2D}(k)$ shown in Fig.~\ref{fig:5}
is reminescent of the horizontal spectrum (of meridional and zonal
winds) observed in the upper troposphere by the Global Atmospheric
Sampling Program \cite{nastrom1985climatology}
where a transition from a $k^{-3}$ to a
$k^{-5/3}$ spectrum at small scales is observed. 
We remark that, despite the similarities between the two spectra,
the physical mechanisms are probably different (see, for example,
\cite{lilly1983stratified}, \cite{vallgren2011possible}) as the transition
in the atmospheric spectrum is observed at a scale around $500 \, km$,
while in our simulations it occours at a scale comparable with
tickness of the layer. 

\subsection{Spectral fluxes}

\begin{figure}[h!]
\centering
\includegraphics[width=1.0\columnwidth]{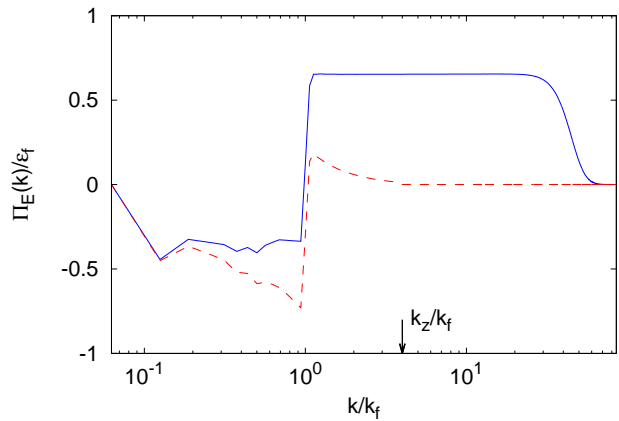}
\caption{Spectral energy flux $\Pi_E(k)$ (blue solid line) 
and 2D energy flux (red, dashed line),  
at $t = 100 \tau_f$}
\label{fig:6}
\end{figure}
The analysis of the spectral flux of the total kinetic energy 
$\Pi_E(k)$ shows that 
in a thin fluid layer the energy injected by the forcing at the
wavenumber $k_f$ indeed splits in two parts (see Figure~\ref{fig:6}).   
At low wavenumbers $k < k_f$ we observe a constant, negative flux of
energy, which indicates the presence of an inverse energy transfer toward large
scales. At high wavenumbers $k > k_f$ we also observe a constant energy flux,
now positive, indicating a direct cascade of energy toward small scales. 
It is worth to remind 
that the kinetic energy transported in the direct cascade 
is not only that of the vertical component of the velocity, 
but contains also the contributions of the modes $k_3 \neq 0$ 
of the horizontal velocities. In this range of scales the dynamics 
in the vertical and horizontal directions is strongly coupled, 
and the positive energy flux cannot be explained 
in terms of a direct cascade of the vertical velocity 
passively transported by a two-dimensional, three components (2D3C) 
flow~\cite{campagne2014direct,moffatt2014note,biferale2017from}.

As we have shown in Fig.~\ref{fig:4}, the inverse cascade involves 
mainly the energy of the 2D mode. 
This observation suggests to check whether or not this inverse cascade 
coincides with a pure 2D dynamics of the 2D mode ${\bm u}^{2D}$. 
To this purpose we have taken the fields ${\bm u}^{2D}$ and 
we have truncated them at $k_h = k_z$ by setting to zero all the modes
with $k_h > k_z$. Then, we have computed the 2D spectral fluxes of the
truncated 2D fields, assuming that they were solutions of the
two-dimensional Navier-Stokes equations. Surprisingly, we find that the
2D energy flux in the range of scales of the inverse cascade does not
coincide with the 3D flux shown in Fig.~\ref{fig:5}.
The physical interpretation of this result
is that the energy of the 2D mode is not simply transported 
toward large scales by the 2D flow itself, but the 3D modes contribute 
to the transport process.
This contrasts with a 2D3C scenario at large scales, 
with the vertical velocity being passively transported by the 2D flow
~\cite{campagne2014direct,moffatt2014note,biferale2017from}.

\begin{figure}[h!]
\centering
\includegraphics[width=1.0\columnwidth]{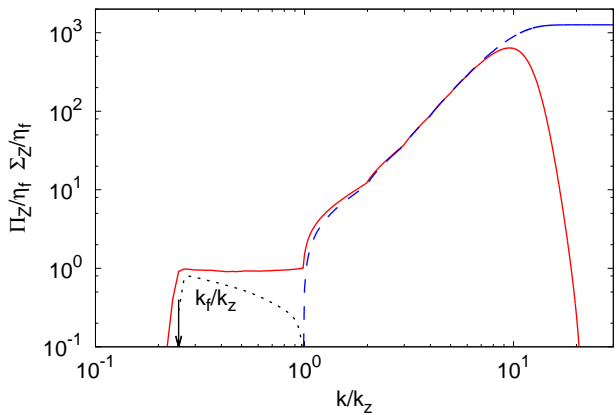}
\caption{Spectral enstrophy flux $\Pi_Z(k)$ (red solid line), 
Spectral enstrophy production $\Sigma_Z(k)$ (blue, dashed line), 
and 2D enstrophy flux (black, dotted line),  
at $t = 100 \tau_f$}
\label{fig:7}
\end{figure}
As discussed in Section~\ref{sec:2}, the main difference between 3D
and 2D Navier-Stokes equations is the absence of the vortex stretching
term ${\bm \omega} \cdot {\bm \nabla} {\bm u}$ in the latter. 
The presence of two positive-defined inviscid invariants (energy
and enstrophy) causes the reversal of the direction of the energy
cascade. Even if the enstrophy is not conserved by the 3D dynamics, 
it is tempting to conjecture that the development of the inverse
cascade in thin fluid layers is due to a dynamical suppression of 
the enstrophy production. 
To investigate this issue we computed the 
total spectral enstrophy flux $\Pi_z(k)$ and the 
total spectral enstrophy production $\Sigma_Z(k)$ 
defined as: 
\begin{eqnarray}
\Pi_{Z}=\int_{|\boldsymbol{q}|\leq{k}}d\boldsymbol{q}(\boldsymbol{v}
\cdot\nabla\boldsymbol{\omega})(\boldsymbol{q})\boldsymbol{\omega}^{\ast}(\boldsymbol{q}) \, ,
\\
\Sigma_{Z}=\int_{|\boldsymbol{q}|\leq{k}}d\boldsymbol{q}(\boldsymbol{\omega}
\cdot\nabla\boldsymbol{v})(\boldsymbol{q})\boldsymbol{\omega}^{\ast}(\boldsymbol{q}) \, .
\end{eqnarray}

In the range of wavenumbers $k_f < k < k_z$ the 
production of enstrophy is negligible, and the enstrophy flux is
constant, as shown in Fig.~\ref{fig:7}. 
This constant flux corresponds to the presence of a direct enstrophy cascade. 
At high wavenumbers, $k > k_z$, the enstrophy production becomes significant
and therefore the enstrophy flux is not constant anymore but grows
following the production term. 
Our results show that,  
in analogy with the case of ideal 2D Navier-Stokes
equations, also in a thin fluid layer the emergence of the
inverse energy cascade is due to the presence of a ``quasi
invariant'', the enstrophy, which is conserved by the
large-scale dynamics. 
Nonetheless, it is worth to notice, that the conservation of the enstrophy 
is not due to the transport by the 2D mode itself. 
Following the same procedure described in the case of the energy flux,
we have computed the 2D enstrophy flux from the truncated 2D velocity
fields. In the range of scale $k_f < k < k_z$ the 2D flux is positive,
but not constant, thus indicating that the full 3D dynamics is
required for the conservation of enstrophy (see Fig.~\ref{fig:7}).

The presence of an intermediate direct enstrophy 
cascade in the range of scales $L_z < \ell < L_f$ allows 
to derive a simple dimensional argument for the scaling of the 
energy flux toward small scales. 
A direct cascade with constant enstrophy flux $\eta_f = \varepsilon_f k_f^2$ 
carries also a residual energy flux, which decreases as 
$\Pi(k) \sim \varepsilon_f (k_f/k)^2$. 
By assuming that the flux of energy of the 
direct cascade which starts from $k_z$ 
is equal to the residual flux $\Pi(k_z)$
carried by the enstropy cascade at the scale $k_z$, 
one gets the prediction 
$\varepsilon_\nu = \Pi(k_z) \sim \varepsilon_f S^2$. 
This prediction has been verified in shell models for quasi-two-dimensional turbulence
\cite{boffetta2011shell}.

\section{Conclusions}
\label{sec:4}
In this paper we present a numerical study 
of the phenomenology of a turbulent flow confined in a thin fluid layer. 
We discuss the possibility to disentangle the complex mixture of 2D and 3D
dynamics by a suitable decomposition of the velocity field in 2D and 3D modes. 

In analogy with previous studies 
\cite{smith1996crossover,celani2010turbulence}, 
when the flow is forced at scales $L_f$ 
larger that the thickness $L_z$ of the layer 
we observe a splitting of the energy cascade in two
directions. A fraction of the energy is transported toward large
scale, giving rise to an inverse energy cascade, while the remnant
energy is transported toward the small viscous scales, as in 3D
turbulence. 
We show that the inverse energy cascade is accompanied by 
the development of a direct cascade of enstrophy in the 
intermediate range of scales $L_z < \ell < L_f$. 
The enstrophy production becomes relevant 
only at small scales $\ell < L_z$, 
allowing for a partial conservation of the enstrophy 
by the large-scale dynamics. 

The scenario which emerges from our findings 
is a coexistence of 2D phenomenology, with a
double cascade of energy and enstrophy {\it \`a la Kraichnan} 
at large scales $\ell > L_z$ 
and a 3D direct energy cascade {\it \`a la Kolmogorov} 
at small scales $\ell < L_z$. 
Interestingly,  
the decomposition of the velocity field in the 2D modes 
and the remaining 3D part 
reveals that the 2D and 3D dynamics are deeply entangled. 
On one hand, we find that the energy and enstrophy 
which are involved in the double cascade at large scales 
are those of the 2D modes. 
On the other hand, the 3D velocity 
is necessary to guarantee a constant flux of 
2D energy and enstrophy in the large-scale transport. 

We plan to extend the analysis 
of the interactions between 2D and 3D modes
to the case of rotating and stably stratified thin fluid layers. 
Previous results~\cite{deusebio2014dimensional} shows that 
rotation causes a suppression of the enstrophy production 
similar to the effects of confinement, 
favoring the two-dimensionalization of the flow 
and the development of the inverse energy cascade.
Nonetheless, this effect is not accompanied by the presence
of a range of scales in which the enstrophy is conserved by the large-scale dynamics. 
This is likely to affect the interactions between 2D and 3D modes. 
In the case of stably stratified fluid layers, 
it has been shown that the conversion of kinetic energy
into potential energy, which is promptly transferred toward the small diffusive scales,
provides a fast dissipative mechanism which suppresses 
the large scale energy transfer~\cite{sozza2015dimensional}. 
Investigating the interactions between 
2D vortical modes and 3D potential modes 
will improve the understanding of this process.

\begin{acknowledgments}
Simulations have been performed at the
Juelich Forschungszentrum (Germany) within the
PRACE Prepatory project PRPA22 and 
at CSC (Finland) within the European project 
HPC-Europa2 ``Energy transfer in turbulent fluid layers''.

We acknowledge useful discussions with A. Celani 
and P. Muratore Ginanneschi and the support 
by the European COST Action MP1305 ``Flowing Matter''.
\end{acknowledgments}

\appendix
\section{2D model for thin layers}
\label{appA}

In this Appendix we discuss a two-dimensional 
model to describe the dynamics of a thin layer
in which the both the thickness $L_z$ 
and the Kolmogorov scale $L_\nu$ tend to zero,
but their ratio remains of order unity.

We consider Navier-Stokes equations~(\ref{eq:1})
in a box in which the horizontal dimensions $L_x=L_y=L$
are much larger than the vertical thickness $L_z = \epsilon L$.
The aspect ratio of the box is determined by the ratio $\epsilon = L_x/L_z$. 
We assume periodic b.c. in all the directions. 
We assume also that the external force ${\bm f}$ 
acts only on horizontal components and depends only 
on horizontal coordinates: ${\bm f}({\bm x})=(f_x(x,y),f_y(x,y),0)$.

The assumption that the thickness $L_z$ 
of the layer is very small and it is of the order 
of the viscous scale, allows to suppose that the modes $k_3 > k_z$ 
are suppressed by the viscosity, and can be neglected. 
Therefore we make a Fourier truncation in the vertical direction 
by retaining only the first modes in the vertical direction
$k_3 = 0, \pm k_z$ where $k_z=2\pi/L_z = O(\epsilon^{-1})$. 
The velocity fields can be expanded as 
\begin{eqnarray}
u_l &=& u^0_l + \sqrt{2}\left[ u_l^c \cos(k_zz) + u_l^s \sin(k_zz) \right] 
\; \\
u_3 &=& u^0_3 + \sqrt{2}\left[ u_3^c \cos(k_zz) + u_3^s \sin(k_zz) \right] \; .  
\end{eqnarray} 
where $l \in [1,2]$. 
Within this notation, 
the 2D mode ${\bm u}^{2D}$ 
introduced in Sec.~\ref{sec:2} coincides with the 
field ${\bm u}^0 = (u_1^0, u_2^0)$. 

The incompressibility condition $\nabla \cdot {\bm u} = 0$ 
gives: 
\begin{equation}
\label{eq:incompr} 
\partial_l u_l^0 = 0 \; ; \; 
\partial_l u_l^c = - k_z u_3^s\; ; \;
\partial_l u_l^s = k_z u_3^c \;. 
\end{equation}
This shows that the 2D mode ${\bm u}^0$ 
satisfies the 2D incompressibility.
The fields $u_3^{s,c}$ are determined by the 
compressibility of the fields ${\bm u}_l^{s,c}$.
Expanding at leading order in $\epsilon$ 
the Navier-Stokes equations~(\ref{eq:1}) 
one obtains the equations
for the fields $u_l^0$,$u_l^{s,c}$ and $u_3^0$: 
\begin{eqnarray}
\label{eq:model1}
  \partial_t u^0_l + u^0_n \partial_n u^0_l &=&  
- \partial_l p + \nu \partial_n \partial_n u^0_l + f_l
\\ \nonumber 
&& - \partial_n (u^c_n u^c_l + u^s_n u^s_l ) 
\\
\label{eq:model2}
  \partial_t u^c_l + u^0_n \partial_n u^c_l &=&  
  \nu \partial_n \partial_n u^c_l - \alpha u^c_l 
\\ \nonumber 
&& - u^c_n \partial_n u^0_l + u^0_3 k_z u^s_l 
\\
\label{eq:model3}
  \partial_t u^s_l + u^0_n \partial_n u^s_l &=&  
  \nu \partial_n \partial_n u^s_l - \alpha u^s_l 
\\ \nonumber 
&& - u^s_n \partial_n u^0_l - u^0_3 k_z u^c_l 
\\
\label{eq:model4}
  \partial_t u^0_3 + u^0_n \partial_n u^0_3 &=&
  \nu \partial_n \partial_n u^0_3  
\\ \nonumber 
&& + \left( u^c_n \partial_n\partial_l u^s_l - u^s_n  \partial_n\partial_l u^c_l
\right)/k_z
\; ,  
\end{eqnarray} 
where $l,n \in [1,2]$ 
and summation over repeated indices is assumed. 

The linear friction term $-\alpha {\bm u}^{s,c}$ in the 2D model 
comes from the derivatives in the vertical direction of the viscous 
term in Eq.~(\ref{eq:1}). 
At finite viscosity $\nu$, in the limit $k_z \to \infty$ 
the friction coefficient $\alpha = \nu k_z^2$ diverges.
The modes ${\bm u}^{c,s}$ are therefore exponentially suppressed. 
In this limit, the equation for the 2D mode ${\bm u}^0$ reduces to 
the two-dimensional Navier-Stokes equations. 

It is interesting also to consider the limit $k_z \to \infty$, 
in which the viscosity vanishes as $\nu \sim k_z^{-2} \to 0$ 
such that $\alpha =  \nu k_z^2 \to const$. 
In this case, the 2D mode ${\bm u}^0$ 
remains coupled with the 3D modes ${\bm u}^{s,c}$.  
The latters are transported and stretched 
by the gradients of the velocity field 
${\bm u}^0$ and are coupled among themselves 
as an harmonic oscillator 
whose frequency is determined by the scalar field $u^0_3$. 
 
The stretching of the fields ${\bm u}^{s,c}$ 
is contrasted by the linear relaxation term $-\alpha {\bm u}^{s,c}$. 
If the dissipation prevails the two fields are exponentially damped, 
and their feedback on the 2D velocity field ${\bm u}^0$ 
can be neglected. 
Conversely, when stretching dominates, 
part of the kinetic energy is transferred 
to the fields ${\bm u}^{s,c}$  
and it is dissipated by the friction.
The transition between the two regimes is expected to occur when 
$\lambda \sim \alpha$, where $\lambda$ is a suitable measure 
of the intensity of the gradients of the 2D mode. 
A dimensional estimate based on the scaling laws 
of the direct enstrophy cascade gives 
$\lambda \sim \eta_f^{1/3}$ where $\eta_f$ is the enstrophy flux. 
Recalling that the viscous scale $L_{\nu}$ 
(defined as the scale at which $Re=1$ ) 
is $L_{\nu} = \nu^{1/2} \eta_f^{-1/6}$ 
and that $\alpha \sim \nu L_z^{-2}$, 
the condition $\lambda > \alpha$ for the transition 
is equivalent to $L_z > L_{\nu}$. 
This shows that 3D modes can be excited only 
when the thickness $L_z$ is larger than the viscous scale $L_\nu$. 

The development of the 3D modes causes a reduction 
of the growth rate of the 2D mode. 
This can be seen from the energy balance of the 2D model: 
\begin{equation}
\label{eq:balance_model}
\frac{d}{dt}(E^{2D}+E^{3D}) = \varepsilon_f - 2\alpha E^{3D}
\end{equation} 
where 
$E^{2D} = 1/2 \langle |{\bm u}^0|^2 \rangle$ 
is the energy of the 2D mode and 
$E^{3D} = 1/2 \langle |{\bm u}^c|^2 + |{\bm u}^s|^2 \rangle$ 
is the energy of the 3D modes. 
For $L_z > L_\nu$ the $E^{3D}$ attains a positive, 
statistically constant value, 
and therefore the growth rate $dE^{2D}/dt$ reduces 
to $\varepsilon_\alpha = \varepsilon_f - 2\alpha E^{3D} < \varepsilon_f$. 

We notice that the model does not provide
a quantitative estimate of the energy $E^{3D}$ 
which appears in Eq.~\ref{eq:balance_model},
and therefore it is not possible to determine
the scaling dependence of the growth 
rate of $E^{2D}$ on the ratio $L_z / L_\nu$.


%

\end{document}